\documentclass[twocolumn,showpacs,preprintnumbers,pra,aps,amssymb,amsfonts,amstex]{revtex4}

\usepackage{times,graphicx,bbm}

\newcommand{\trace}{\mathop{\rm Tr}\nolimits}

\newcommand{\diag}{\mathop{\rm Diag}\nolimits}

\newcommand{\bra}[1]{\langle#1|}
\newcommand{\ket}[1]{|#1\rangle}

\newcommand{\qed}{\hfill$\square$\par\vskip24pt}
\newcommand{\cB}{{\cal B}}

\newcommand{\cH}{{\cal H}}

\newcommand{\cL}{{\cal L}}

\newcommand{\C}{{\mathbb{C}}}

\newcommand{\identity}{\mathbbm{1}}

\newtheorem{theorem}{Theorem}
\newtheorem{lemma}{Lemma}

\newtheorem{proposition}{Proposition}

\begin{document}
\title{Continuity bounds on the quantum relative entropy}
\author{Koenraad M.R. Audenaert} 
\email{k.audenaert@imperial.ac.uk}
\author{Jens Eisert}
\email{jense@imperial.ac.uk}
\affiliation{Blackett Laboratory, Imperial College London,
Prince Consort Rd, London SW7 2BW, UK}
\affiliation{Institute for Mathematical Sciences, Imperial College London,
Exhibition Rd, London SW7 2BW, UK}
%\affiliation{3 Universit{\"a}t Potsdam,
%Institut f{\"u}r Physik, Am Neuen Palais 10,
%D-14469 Potsdam, Germany}

\date{\today}

%------------------------------------------------------------------ ABSTRACT
\begin{abstract}
The quantum relative entropy 
is frequently used as a distance, or distinguishability measure
between two quantum states.
In this paper we study the relation between this measure and a 
number of other measures used for that purpose,
including the trace norm distance.
More specifically, 
we derive lower and upper bounds on the relative entropy in terms of
various distance measures for the difference of the states 
based on unitarily invariant norms.
The upper bounds can be considered as statements of continuity
of the relative entropy distance in the sense of Fannes.
We employ methods from optimisation theory 
to obtain bounds that are as sharp as possible.

\end{abstract}
\maketitle
%------------------------------------------------------------------ BODY

%\begin{multicols}{2}

\section{Introduction}
The relative entropy of states of quantum systems
is a measure of how well one quantum state can be operationally
distinguished from another. Defined as
\begin{eqnarray}\nonumber
    S(\rho||\sigma)=
    \trace[\rho(\log \rho  - \log \sigma) ]
\end{eqnarray}
for states $\rho$ and $\sigma$, it quantifies the extent to which one
hypothesis $\rho$ differs from an alternative hypothesis $\sigma$
in the sense of quantum hypothesis testing
\cite{ohya,Wehrl,disti,disti2,disti3}.
Dating back to work by Umegaki \cite{umegaki}, the relative entropy is a
quantum generalisation
of the Kullback-Leibler relative entropy for probability
distributions in mathematical statistics \cite{Kullback}.
The quantum relative entropy plays an
important role in quantum statistical mechanics \cite{Wehrl}
and in quantum information
theory, where it appears as a central notion in the
study of capacities of quantum channels
\cite{Schumacher,Holevo,Schumacher2,prop}
and in entanglement theory \cite{prop,Plenio,Plenio2}.

In {\it finite-dimensional Hilbert spaces}, 
the relative entropy functional
is manifestly continuous \cite{Wehrl}, 
see also footnotes \footnote{
    For states on infinite-dimensional Hilbert spaces
    the relative entropy functional is not trace norm 
    continuous any more,
    but --  as the von-Neumann entropy -- lower semi-continuous. That is, 
    for  sequences of states $\{\sigma_n\}_n$  and 
	$\{\rho_n\}_n$	
converging in trace norm to states $\sigma$ and $\rho$, i.e.,
$\lim_{n\rightarrow\infty} \| \sigma_n-\sigma\|_1 =0$
and $\lim_{n\rightarrow\infty} \| \rho_n-\rho\|_1 =0$,
we merely have 
$	S(\rho||\sigma)\leq \liminf_{n\rightarrow\infty}
	S(\rho_n||\sigma_n)$.
    However, for systems for which the Gibbs state exists, these
discontinuities can be tamed \cite{Wehrl} when considering compact
subsets of state space with finite mean energy.
    In a similar manner, entropic
    measures of entanglement can become
    trace norm continuous on
    subsets with bounded energy \cite{Jensito}.}, \footnote{
    For considerations of the continuity of the relative entropy
    in classical contexts, see Ref.\ \cite{naudts}.}.
In particular, if 
	$\{\sigma_n\}_{n}$ is a sequence of states
	of fixed
    finite dimension satisfying 
\begin{equation}\nonumber
	\lim_{n\rightarrow \infty} || \sigma_n - \sigma ||_1 =
	\lim_{n\rightarrow \infty} \trace | \sigma_n - \sigma |  
	=0
\end{equation}
for a given state $\sigma$, then
\begin{eqnarray}\nonumber
    \lim_{n\rightarrow\infty} S(\sigma_n|| \sigma)=0.
\end{eqnarray}
In practical contexts, however, more precise estimates can be
necessary, in particular in an asymptotic setting.
Consider a state $\rho$ on a Hilbert space $\cH$, and a sequence 
$\{\sigma_n\}_n$,
where $\sigma_n$ is a state on $\cH^{\otimes n}$, the $n$-fold tensor product of $\cH$.
The sequence is said to asymptotically approximate $\rho$ if $\sigma_n$ tends to $\rho^{\otimes n}$ for
$n\rightarrow \infty$. 
More precisely, one typically requires that 
\begin{equation}\nonumber
	\lim_{n\rightarrow\infty}
	\| \sigma_n - \rho^{\otimes n}\|_1 =0.
\end{equation}
Now, as an alternative to the trace norm distance, 
one can consider the use of the Bures distance.
The Bures distance $D$ is defined as
$$
D(\rho_1,\rho_2) = 2\left(1-F(\rho_1,\rho_2)\right)^{1/2},
$$
in terms of the Uhlmann fidelity
$$
F(\rho_1,\rho_2) = \trace( \rho_1^{1/2} \rho_2 \rho_1^{1/2})^{1/2}.
$$
Because of the inequalities \cite{Hayden}
\begin{equation}\label{hayden}
1-F(\rho_1,\rho_2) \le \trace|\rho_1-\rho_2| \le 
\left(1-F^2(\rho_1,\rho_2)\right)^{1/2},
\end{equation}
the trace norm distance tends to zero if and only if the Bures distance tends to zero, which shows that,
for the purpose of state discrimination, both distances are essentially equivalent and one can use whichever
is most convenient.

A natural question that now immediately arises 
is whether the same statement is true for the relative entropy.
To find an answer to that one would need inequalities like (\ref{hayden}) connecting
the quantum relative entropy, used as a distance measure, 
to the trace norm distance, or similar distance measures.

In this paper, we do just that: we find upper bounds on the relative entropy functional
in terms of various norm differences of the two states. As such, the presented
bounds are very much in the same spirit as Fannes' inequality, sharpening the
notion of continuity for the von Neumann entropy \cite{fannes}.
It has already to be noted here that one of the main stumbling blocks in this undertaking
is the well-known fact that the relative entropy is not a very good distance measure,
as it gives infinite distance between non-identical pure states. However, we will present
a satisfactory solution, based on using the minimal eigenvalue of the state
that is the second argument of the relative entropy.
Apart from the topic of upper bounds, we also study lower bounds on the relative entropy, giving a
complete picture of the relation between norm based distances and relative entropy.

We start in Section II
with presenting  a short motivation of how this paper came about.
Section III contains the relevant notations, definitions and basic results that will be used
in the rest of the paper. In Section IV we discuss some properties of unitarily invariant (UI) norms
that will allow us to consider all UI norms in one go.
The first upper bounds on the relative entropy $S(\rho||\sigma)$ are presented in Section V, one bound being quadratic
in the trace norm distance of $\rho$ and $\sigma$ and the other logarithmic in the minimal eigenvalue of
$\sigma$. Both bounds separately capture an essential behaviour of the relative entropy, and it is
argued that finding a single bound that captures both behaviours at once is not a trivial undertaking.
Nevertheless, we will succeed in doing this in Section VII by constructing upper bounds that are as sharp
as possible for given trace norm distance \textit{and} minimal eigenvalue of $\sigma$.
In Section VI we use similar techniques to derive lower bounds that are as sharp as possible.
Finally, in Section IX, we come back to the issue of state discrimination mentioned at the beginning.
%%%%%%%%%%%%%%%%%%%%%%%%%%%%%%%%%%%%%%%%%%%%%%%%%%%%%%%%%%%%%%%%%%%%%%%%%%%%%%%%%%%%%%%%

\section{Background}

In Ref.\ \cite{brat2}
(Example 6.2.31, p.\ 279) we find the following upper bound on the relative entropy, valid
for all $\rho$ and for non-singular $\sigma$:
\begin{equation}\label{bound_brat}
S(\rho||\sigma) \le \frac{||\rho-\sigma||_\infty}{\lambda_{\min}(\sigma)}.
\end{equation}
This bound is linear in the operator norm distance between $\rho$ and $\sigma$
and has a $1/x$ dependence on $\lambda_{\min}(\sigma)$.
For several purposes, such a bound is not necessarily sharp enough.
The impetus for the present paper was given by the observation that
sharper upper bounds on the relative entropy should be possible than (\ref{bound_brat}).
Specifically, there should exist bounds that are
\begin{enumerate}
\item {\em quadratic} in $\rho-\sigma$, and/or
\item depend on $\lambda_{\min}(\sigma)$ in a {\em logarithmic} way.
\end{enumerate}
A simple argument shows that a logarithmic dependence on $\lambda_{\min}(\sigma)$ can be
achieved instead of an $1/x$ dependence.
Note that $0\ge\log\sigma\ge\identity\cdot\log\lambda_{\min}(\sigma)$.
Thus,
\begin{eqnarray}
S(\rho||\sigma) &=& \trace[\rho(\log\rho-\log\sigma)] \nonumber \\
&\le& -S(\rho)-\log\lambda_{\min}(\sigma) \nonumber \\
&\le& -\log\lambda_{\min}(\sigma). \label{bound1}
\end{eqnarray}
Concerning the quadratic dependence on $\rho-\sigma$, we can put $\rho=\sigma+\varepsilon\Delta$,
with $\trace[\Delta]=0$, and calculate the derivative
$$\lim_{\varepsilon\rightarrow0} S(\sigma+\varepsilon\Delta||\sigma) / \varepsilon
$$
and find that this turns out to be zero for any non-singular $\sigma$.
Indeed, the gradient of the relative entropy $S(\rho||\sigma)$ with respect to $\rho$ is
$\identity+\log\rho-\log\sigma$ (see Lemma \ref{lemma1}). 
Hence, for $\rho=\sigma$ and 
$\trace[\Delta]=0$,
\begin{eqnarray}\nonumber
    \lim_{\varepsilon\rightarrow0} S(\sigma+\varepsilon\Delta||\sigma)/\varepsilon
    =\trace[\Delta(\identity+\log\sigma-\log\sigma)]=0.
\end{eqnarray}
This seems to imply that for small $\varepsilon$, $S(\sigma+\varepsilon\Delta||\sigma)$ must at least be
quadratic in $\varepsilon$, and, therefore, upper bounds might exist that indeed are quadratic in $\varepsilon$.
Furthermore, Ref.\ \cite{ohya} contains the following quadratic lower bound 
(Th.\ 1.15)
\begin{equation}\nonumber
\label{bound_ohya}
S(\rho||\sigma) \ge \frac{1}{2}||\rho-\sigma||_1^2.
\end{equation}
The rest of the paper will be devoted to finding firm evidence for these intuitions, by exploring
the relation between relative entropy and norm based distances, culminating in a number of bounds
that are the sharpest possible.
%%%%%%%%%%%%%%%%%%%%%%%%%%%%%%%%%%%%%%%%%%%%%%%%%%

\section{Notation}

In this paper, we will use the following notations.
We will use the standard vector and matrix bases:
$e^i$ is the vector with the $i$-th element equal to 1, and all other elements 
being equal to $0$.
$e^{i,j}$ is the matrix with $i,j$ element equal to 1 and all other elements 0.
For any diagonal matrix $A$, we write $A_i$ as a shorthand for $A_{i,i}$,
and $\diag(a_1,a_2,\ldots)$ is the diagonal matrix with $a_i$ as diagonal elements.
We reserve two symbols for the following special matrices:
\begin{equation}\nonumber
E:=\diag(1,0,\ldots,0) = e^{1,1},
\end{equation}
and
\begin{equation}\nonumber
F:=\diag(1,-1,0,\ldots,0) = e^{1,1}-e^{2,2}.
\end{equation}
The positive semi-definite order is denoted using the $\ge$ sign: $A\ge B$ iff 
$A-B\ge 0$ (positive semi-definite).

The (quantum) relative entropy is denoted as 
$S(\rho||\sigma) = \trace[\rho(\log\rho-\log\sigma)]$.
All logarithms in this paper are natural logarithms.
When $\rho$ and $\sigma$ are both diagonal (i.e., when we encounter the 
commutative, classical case) we use the shorthand
\begin{eqnarray*}
\lefteqn{S((r_1,r_2,\ldots)||(s_1,s_2,\ldots))} \\
&:=& S(\diag(r_1,r_2,\ldots)||\diag(s_1,s_2,\ldots)).
\end{eqnarray*}
\begin{lemma}\label{lemma1}
The gradient of the relative entropy $S(\rho||\sigma)$ with respect to its first argument $\rho$, being non-singular, is given by
	$\identity+\log\rho-\log\sigma$.
\end{lemma}

\textit{Proof.}
The calculation of this derivative is straightforward.
%\begin{eqnarray*}
%	\lim_{\varepsilon\rightarrow0} \left(
%	S(\rho+\varepsilon\Delta||\sigma)-
%	S(\rho||\sigma)\right)
%\varepsilon &=&
%	\frac{\partial}{\partial\varepsilon}\Big|_{\varepsilon=0} 
%	S(\rho+\varepsilon\Delta||\sigma) 
%\end{eqnarray*}
Since the classical entropy function 
$x\longmapsto h(x) := -x \log x $ is continuously differentiable
on $(0,1)$, and therefore,
\begin{eqnarray*}
	\frac{\partial}{\partial\varepsilon}\Big|_{\varepsilon=0} 
	S(\rho+\varepsilon \Delta  ) =
	\trace[\Delta h'(\rho) ],
\end{eqnarray*}
we can write
\begin{eqnarray}
\label{log_deriv}
   \lim_{\varepsilon\rightarrow0} S(\rho+\varepsilon\Delta||\sigma)/\varepsilon
  &=&  \trace[\Delta(\identity+\log\rho-\log\sigma)].\nonumber
\end{eqnarray}
%\newpage
%Using the 
%integral formulation of the Fr\'echet derivative of the matrix logarithm,
%\begin{equation}
%\label{log_deriv}
%\frac{\partial}{\partial\varepsilon}\Big|_{\varepsilon=0} \log(\rho+
%\varepsilon\Delta) =
%\int_0^\infty dx (\rho+x)^{-1} \Delta (\rho+x)^{-1},
%\end{equation}
%the derivative of the von Neumann entropy is
%\begin{eqnarray*}
%&&\frac{\partial}{\partial\varepsilon}\Big|_{\varepsilon=0} S(\rho+%
%\varepsilon\Delta) \\
%&=&-\trace[\Delta\log\rho] - \trace\left[\rho\int_0^\infty dx (\rho+x)^{-1} 
%\Delta (\rho+x)^{-1}\right] \\
%&=&-\trace[\Delta\log\rho] - \trace\left[\int_0^\infty dx(\rho+x)^{-1}\rho 
%(\rho+x)^{-1} \Delta \right] \\
%&=&-\trace[\Delta\log\rho] - \trace[\Delta],
%\end{eqnarray*}
%for all non-singular $\rho$, such that then
%\begin{eqnarray} \label{relent_deriv}
%    \lim_{\varepsilon\rightarrow0} S(\rho+\varepsilon\Delta||\sigma)/%
%\varepsilon
%    =\trace[\Delta(\identity+\log\rho-\log\sigma)].
%\end{eqnarray}
\qed

Finally, we recall a number of series expansions related to the logarithm, which are valid for
$-1<y<1$,
\begin{eqnarray}
\log(1-y) &=& -\sum_{k=1}^\infty \frac{y^k}{k},\nonumber\\
\log(1+y)+\log(1-y) &=& -\sum_{k=1}^\infty\frac{y^{2k}}{k} ,\nonumber\\
\log(1+y)-\log(1-y) &=& 2\sum_{k=0}^\infty\frac{y^{2k+1}}{2k+1}.
\nonumber
\end{eqnarray}
These expansions will be made extensive use of.

%%%%%%%%%%%%%%%%%%%%%%%%%%%%%%%%%%%%%%%%%%%%%%%%%
\section{Unitarily Invariant Norms}
In this section we collect the main definitions and known results about unitarily invariant norms
along with a number of refinements that will prove to be very useful for the rest of the paper.

A \textit{unitarily invariant norm} (UI norm), denoted with
$|||.|||$, is a norm on square matrices
that satisfies the property 
\begin{equation}\nonumber
|||UAV|||=|||A|||
\end{equation}
for all $A$ and for unitary $U$, $V$ (\cite{bhatia}, Section IV.2).
Perhaps the most important property of UI norms is that they only depend on the singular values of the matrix $A$.
If $A$ is positive semi-definite, then $|||A|||$ depends only on the eigenvalues of $A$.

A very important class of UI norms are the \textit{Ky Fan norms} $||.||_{(k)}$, which are defined as follows:
for any given square $n\times n$ matrix $A$, with singular values $s_j^\downarrow(A)$ (sorted in non-increasing order)
and $1\le k\le n$, the $k$-Ky Fan norm is the sum of the $k$ largest singular values of $A$:
$$
||A||_{(k)} = \sum_{j=1}^k s_j^\downarrow(A).
$$
Two special Ky Fan norms are the \textit{operator norm} 
and the \textit{trace norm},
\begin{equation}\nonumber
||A||_\infty = ||A||_{(1)},\,\,
||A||_{\mbox{Tr}} = ||A||_1 = ||A||_{(n)}.
\end{equation}
The importance of the Ky Fan norms derives from their leading role in Ky Fan's \textit{Dominance Theorem}
(Ref.\ \cite{bhatia}, Theorem IV.2.2):
\begin{theorem}[Ky Fan Dominance]
Let $A$ and $B$ be any two $n\times n$ matrices. If $B$ majorises $A$ in all the Ky Fan norms, 
$$
||A||_{(k)}\le ||B||_{(k)}, 
$$
for all $k=1,2,...$ , 
then it does so in
all other UI norms as well,
$$
  |||A|||\le|||B|||.
$$
\end{theorem}
From Ky Fan's Dominance Theorem follows the following well-known
norm dominance statement.
\begin{lemma}
For any matrix $A$, and any unitarily invariant norm $|||.|||$,
$$
||A||_\infty \le \frac{|||A|||}{|||E|||} \le ||A||_1.
$$
\end{lemma}
\textit{Proof.}
We need to show that, for every $A$,
$$
|||(||A||_\infty)E||| \le |||A||| \le |||(||A||_1)E|||,
$$
holds for every unitarily invariant norm.
By Ky Fan's dominance theorem, we only need to show this for the Ky Fan norms.
All the Ky Fan norms of $E$ are 1, and
$$
||A||_\infty =||A||_{(1)} \le ||A||_{(k)} \le ||A||_{(d)}=||A||_1
$$
follows from the definition of the Ky Fan norms.
\qed

The main mathematical object featuring in this paper is not the state, but rather the difference $\Delta$ of
two states, $\Delta: =\rho-\sigma$, and for that object a stronger dominance result obtains.
We first show that the largest norm difference between two states occurs for orthogonal pure states.
Indeed, by convexity of norms, $|||\rho-\sigma|||$ is maximal in pure $\rho$ and $\sigma$.
A simple calculation then reveals that, for any unitarily invariant norm,
$$
|||\,\ket{\psi}\bra{\psi} - \ket{\phi}\bra{\phi}\,||| = \left(
	1-|\langle\psi|\phi\rangle|^2
	\right)^{1/2}
|||F|||.
$$
This achieves its maximal value $|||F|||$ for $\psi$ orthogonal to $\phi$,
showing that it makes sense to normalise a norm distance $|||\rho-\sigma|||$ by division by $|||F|||$.
We will call this a \textit{rescaled} norm.
We now have the following dominance result for rescaled norms of differences of states:
\begin{lemma}\label{lemma:dom}
For any Hermitian $A$, with $\trace[A]=0$,
$$
\frac{||A||_\infty}{||F||_\infty} \le \frac{|||A|||}{|||F|||} \le \frac{||A||_1}{||F||_1}.
$$
\end{lemma}
Note that equality can be obtained for any value of $|||A|||$, by setting $A=cF$.

\textit{Proof.}
We need to show, for all traceless Hermitian $A$, that
\begin{equation}\label{eq:dom}
|||(||A||_\infty)F||| \le |||A||| \le |||(||A||_1/2) F |||
\end{equation}
holds for every unitarily invariant norm.
Again by Ky Fan's dominance theorem, 
we only need to do this for the Ky Fan norms $||.||_{(k)}$.
Since
$$
||F||_{(k)} = \left\{\begin{array}{cc}1,&k=1,\\2,&k>1,\end{array}\right.
$$
and 
\begin{eqnarray*}
	||X||_\infty = ||X||_{(1) }\le ||X||_{(k)} \le ||X||_{(d)}=||X||_1,
\end{eqnarray*}	
the inequalities (\ref{eq:dom}) follow trivially for $k>1$.
The case $k=1$ is covered by Lemma \ref{lemma:k1} below.
\qed

\begin{lemma}\label{lemma:k1}
For any Hermitian $A$, with $\trace[A]=0$,
$$
||A||_1 \ge 2||A||_\infty.
$$
\end{lemma}
\textit{Proof.}
Let the Jordan decomposition of $A$ be 
\begin{equation}\nonumber
	A=A_+ - A_-, 
\end{equation}
with $A_+$, $A_-\ge0$.
Since $\trace[A]=0$, clearly $\trace[A_+]=\trace[A_-]$ holds.
Thus, $||A||_1=\trace|A|=\trace[A_+]+\trace[A_-]=2\trace[A_+]$.
Also, 
\begin{equation}\nonumber
	||A||_\infty = \max(||A_+||_\infty,||A_-||_\infty).
\end{equation}
Hence,
$||A||_\infty \le \max(||A_+||_1,||A_-||_1) = \trace[A_+] = ||A||_1/2$.
\qed
%%%%%%%%%%%%%%%%%%%%

In this paper, we will also be dealing with $\Delta=\rho-\sigma$ under the constraint $\sigma\ge\beta\identity$.
Obviously we have
$$
\beta\le 1/d.
$$
We now show that under this constraint, any rescaled norm of $\Delta$ is upper 
bounded by $1-\beta$.
\begin{lemma}\label{lem:maxt}
For any state $\rho$, and states $\sigma$ such that $\sigma\ge\beta\identity$,
\begin{equation}\nonumber
T:=|||\rho-\sigma|||/|||F||| \le 1-\beta.
\end{equation}
\end{lemma}
\textit{Proof.}
We proceed by maximising $T$ under the constraint $\sigma\ge\beta\identity$.
Convexity of norms yields that $T$ is maximal when $\rho$ and $\sigma$ are extremal
\cite{Convex}, 
hence in $\rho$ being a pure state
$\ket{\phi}\bra{\phi}$ and $\sigma$ being
of the form
\begin{equation}\nonumber
	\sigma = \beta\identity + (1-\beta d)\ket{\psi}\bra{\psi}.
\end{equation}
Fixing $\phi=e^1$, we need to maximise
$$
|||e^{1,1}-\beta\identity-(1-\beta d)\ket{\psi}\bra{\psi}\, |||
$$
over all $\psi$.
Put $\psi=(\cos\alpha,\sin\alpha,0,\ldots,0)^T$, then
the eigenvalues of the matrix are
$$
\lambda_\pm=\left((d-2)\beta\pm \left( {(\beta d)^2+4(1-\beta 
d)\sin^2\alpha} \right)^{1/2} \right)/2
$$
and $-\beta$ (with multiplicity $d-2$).
One finds that, for $d>2$, $\lambda_+,|\lambda_-|\ge\beta$, for any value of $\alpha$,
and both $\lambda_+$ and $|\lambda_-|$ are maximal for $\alpha=\pi/2$, as would be expected.
The maximal Ky Fan norms of this matrix are therefore
\begin{eqnarray*}
||.||_{(1)} &=& \lambda_+ = 1-\beta ,\\
||.||_{(k)} &=& \lambda_+ + |\lambda_-|+(k-2)\beta = (2-\beta d) + (k-2)\beta,
\end{eqnarray*}
for $k>1$.
Hence, for every Ky Fan norm, the maximum norm value is obtained for orthogonal $\phi$ and $\psi$.
By the Ky Fan dominance theorem, this must then hold for any UI norm.
In case of the trace norm, as well as of the operator norm, the rescaled value of the maximum
is $1-\beta$. By Lemma \ref{lemma:dom}, this must then be the maximal value for any rescaled norm.
\qed

\textit{Remark.}
For the Schatten $q$-norm, $|||F||| = 2^{1/q}$.
The largest value of $|||F|||$ is 2, obtained for the trace norm,
and the smallest value is 1, for the operator norm.
%%%%%%%%%%%%%%%%%%%%%%%%%%%%%%%%%%%%%%%%%%%%%%%%%%
\section{Some simple upper bounds}
In this Section we present our first attempts at finding upper bounds that capture the essential features
of relative entropy. In Subsection A we present a bound that is indeed quadratic in the trace norm distance,
the existence of which was already hinted at in Section II.
Likewise, in Subsection B, we find a bound that is logarithmic in the minimal eigenvalue of $\sigma$,
again in accordance with previous intuition.
Combining the two bounds into one that has both of these features turns out to be not so easy.
In fact, in Subsection C a number of arguments are given that initially hinted at the impossibility
of realising sich a combined bound.
Nevertheless, we will succeed in finding a combined bound later on in the paper, by using techniques
from optimisation theory \cite{Convex}.

\subsection{A quadratic upper bound}
\begin{lemma}
For any positive definite matrix $A$ and Hermitian $\Delta$
such that $A+\Delta $ is positive definite,
$$
\log(A+\Delta)-\log(A) \le \int_0^\infty dx(A+x)^{-1}\Delta(A+x)^{-1}.
$$
\end{lemma}
{\em Proof.}
Since the logarithm is strictly matrix concave \cite{HJII},
for all $t\in[0,1]$:
$$
\log((1-t)A+tB) \ge (1-t)\log(A)+t\log(B).
$$
Setting $B=A+\Delta$ and rearranging terms
then gives
$$
\frac{\log(A+t\Delta)-\log(A)}{t} \ge \log(A+\Delta)-\log(A),
$$
for all $t\in[0,1]$. A fortiori,
this holds in the limit for $t$ going to zero, and then the left-hand side is just the
Fr\'echet derivative of $\log$ at $A$ in the direction $\Delta$.
%, given by 
%Eq.\ (\ref{log_deriv}).
\qed

This Lemma allows us to give a simple upper bound on
$S(\sigma+\Delta||\sigma)$. Note that if $A\ge B$ then
$\trace [CA]\ge\trace [CB]$ for any $C\ge 0$. Therefore, we arrive at
\begin{eqnarray*}
S(\rho||\sigma) &=&
\trace\left[(\sigma+\Delta)(\log(\sigma+\Delta)-\log(\sigma))\right] \\
&\le& \int_0^\infty dx\trace[(\sigma+\Delta)(\sigma+x)^{-1}\Delta(\sigma+x)^{-1}] \\
&=& \int_0^\infty dx\trace[(\sigma+x)^{-1}\sigma(\sigma+x)^{-1}\Delta] \\
& + &\int_0^\infty dx\trace[\Delta(\sigma+x)^{-1}\Delta(\sigma+x)^{-1}].
\end{eqnarray*}
The first integral evaluates to $\trace[\Delta]$, because
$$
\int_0^\infty dx \frac{s}{(s+x)^2} = 1
$$ for any $s>0$, and therefore gives the value $0$.
The second integral can be evaluated most easily in a basis in which $\sigma$ is diagonal.
Denoting by $s_i$ the eigenvalues of $\sigma$, we get
\begin{eqnarray}
&&\int_0^\infty dx\trace[\Delta(\sigma+x)^{-1} \Delta(\sigma+x)^{-1}] \nonumber \\
&=& \sum_{i,j} \Delta_{i,j}\Delta_{j,i} \int_0^\infty dx (s_i+x)^{-1}
(s_j+x)^{-1} \nonumber\\
&=& \sum_{i\neq j} \Delta_{i,j}\Delta_{j,i} \frac{\log s_i -\log s_j }{s_i-s_j} +
\sum_i (\Delta_{i,i})^2 \frac{1}{s_i}.\label{dec}
\end{eqnarray}
The coefficients of $\Delta_{i,j}\Delta_{j,i}$ are easily seen to be always positive, and furthermore,
bounded from above by $1/\lambda_{\min}(\sigma)$. Hence we get the upper bound
$$
\int_0^\infty dx\trace[\Delta(\sigma+x)^{-1}\Delta(\sigma+x)^{-1}] \le \frac{\trace[\Delta^2]}{\lambda_{\min}(\sigma)},
$$
yielding an upper bound on the relative entropy which is, indeed, quadratic in $\Delta$:
\begin{theorem}
For states $\rho$ and $\sigma$ with $\Delta=\rho-\sigma$, $T=||\Delta||_2$ and
$\beta=\lambda_{\min}(\sigma)$,
\begin{equation}\nonumber
\label{bound_quad}
S(\rho||\sigma) \le \frac{T^2}{\beta}.
\end{equation}
\end{theorem}
%%%%%%%%%%%%%%%%%%%%%%%%%%%%%%%%%%%%%%%%%%%%%%%%%%%%%%%%%%
\subsection{An upper bound that is logarithmic in the minimum eigenvalue of $\sigma$}
We have already found a sharper bound than (\ref{bound_brat}) concerning its dependence on $\lambda_{\min}(\sigma)$.
However, the bound (\ref{bound1}) is not sharp at all concerning its
dependence on $\rho-\sigma$.
A slight modification can greatly improve this.
First note
\begin{eqnarray*}
|\trace[\Delta\log\sigma]| &\le& ||\Delta||_1 \cdot
||\log\sigma||_{\infty} \\
&=& \trace|\Delta|\cdot|\log\lambda_{\min}(\sigma)|.
\end{eqnarray*}
This inequality can be sharpened, since $\trace[\Delta]=0$ and $\sigma$ is a state.
Let $\Delta=\Delta_+ - \Delta_-$ be the Jordan decomposition of $\Delta$, then
\begin{equation}\nonumber
	|\trace[\Delta\log\sigma]| \le
	||\Delta_+||_1\cdot|\log\lambda_{\min}(\sigma)|, 
\end{equation}
and hence
\begin{eqnarray*}
|\trace[\Delta\log\sigma]| &\le&
\trace|\Delta|/2  \cdot|\log\lambda_{\min}(\sigma)|.
\end{eqnarray*}
Furthermore, we have Fannes' continuity of the von Neumann entropy \cite{fannes},
$$
|S(\sigma+\Delta)-S(\sigma)| \le  T \log d +\min\left(-T\log 
T,\frac{1}{e}\right),
$$
where $d$ is the dimension of the underlying Hilbert space
and $T:=\trace |\Delta| $.
Combining all this with
$$
S(\sigma+\Delta||\sigma) = -(S(\sigma+\Delta)-S(\sigma)) - \trace[\Delta\log\sigma]
$$
gives rise to the subsequent upper bound, logarithmic in the smallest eigenvalue of $\sigma$.
\begin{theorem}
For all states $\rho$ and $\sigma$ on a $d$-dimensional
Hilbert space, with $T=||\rho-\sigma||_1$ and $\beta=\lambda_{\min}(\sigma)$,
\begin{equation}
S(\rho||\sigma) \le T \log d +
\min\biggl (-T\log T,\frac{1}{e}\biggr) - \frac{T\log\beta}{2}.
\label{bound_log}
\end{equation}
\end{theorem}

%%%%%%%%%%%%%%%%%%%%%%%%%%%%%%%%%%%%%%%%%%%%%%%%%%%%%%%%%%%%%%%%%%%%%%%%%%%%
\subsection{A combination of two bounds?}
The following question comes to mind almost automatically:
can we combine the two bounds (\ref{bound_quad}) and (\ref{bound_log})
into a single bound that is both quadratic in $\Delta$
and logarithmic in $\lambda_{\min}(\sigma)$? This would certainly be a very
desirable feature for a good upper bound.
For instance, could it be true that
$$
S(\rho||\sigma) \le C\cdot  \trace[(\rho-\sigma)^2]\cdot |\log\lambda_{\min}(\sigma)|,
$$
for some constant $C>0$?
Unfortunately, the answer to this first attempt is negative.
In fact, the proposed inequality is violated no matter how large the value of $C$.
\begin{proposition}
For any  $r>0$ there exist states $\sigma$ and $\rho$ such that
\begin{equation} 
\label{bound_bad}
S(\rho||\sigma) > r\cdot  \trace[(\rho-\sigma)^2]\cdot |
\log\lambda_{\min}(\sigma)|.
\end{equation}
\end{proposition}
\textit{Proof.}
It suffices to consider the case that $\sigma,\rho$ are states acting on
the Hilbert space $\C^2$, and that $\sigma$ and $\rho$ commute. Hence, the
statement must only be shown for two probability distributions
\begin{eqnarray*}
P=(p,1-p),\,\,\,Q=(q,1-q).
\end{eqnarray*}
Without loss of generality we can require $q$ to be in $[0,1/2]$.
Then, one has to show that for any $r>0$ there exist $p,q$
such that
the $C^\infty$-function $f$,
%$f:[0,1]\times [0,1/2]\times [0,\infty)\rightarrow {\R}$, 
defined as
\begin{eqnarray*}
    f(p,q,r) &=& r \left((p-q)^2 + (2-p-q)^2\right)\, | \log(q)| \\
    &-& \left(
    p \log(p/q) +(1-p) \log[(1-p)/(1-q)]
    \right),
\end{eqnarray*}
assumes a negative value. Now, for any $r>1$, fix a $q\in(0,1/2)$ such that
$- 4 r (q \log q)<1$.
Clearly,
$$
    f(q,q,r) =0,\,\,\,\,\,\,
    \frac{\partial}{\partial p}\bigr|_{p=q} f(p,q,r)=0.
$$
Then
\begin{eqnarray*}
    \frac{\partial^2}{\partial p^2} \bigr|_{p=q} f(p,q,r)
    &=&
    -\frac{1}{1-q} - \frac{1}{q} - 4 r \log(q) \\
    &<&  - \frac{1}{q} - 4 r \log(q)< 0 .
\end{eqnarray*}
This means that there exists an
$\varepsilon>0$ such that $f(p,q,r)<0$ for $p\in[q,q+\varepsilon]$,
which in turn proves the validity of 
(\ref{bound_bad}).
\qed

The underlying reason for this failure is that the two bounds
(\ref{bound_log}) and (\ref{bound_quad}) are incompatible, in the sense that there are two different regimes
where either one or the other dominates.
To see when the logarithmic dependence dominates, let us again take the basis where
$\sigma$ is diagonal, with $s_i$ being the main diagonal elements.
When keeping $\Delta=\rho-\sigma$ fixed while $s_1=\lambda_{\text{min}}(\sigma)$
tends to zero, then
\begin{eqnarray}\nonumber
    \lim_{s_1\rightarrow 0} S(\sigma+\Delta||\sigma)/ |\log s_1| = \Delta_{1,1}<\infty.
\end{eqnarray}
Hence, in the regime where $\lambda_{\min}(\sigma)$ tends to zero and $\rho-\sigma$ is fixed,
the bound (\ref{bound_log}) is the appropriate one.

The other regime is the one where $\sigma$ is fixed and $\rho-\sigma$ tends to zero.
This can be intuitively
seen by considering the case where the states $\rho$ and $\sigma$ commute
(the classical case).
Let $p_i$ and $q_i$ be the diagonal elements of $\rho$ and $\sigma$, respectively,
in a diagonalising basis, and $r_i=p_i-q_i$. Then
\begin{eqnarray}\nonumber
S(\rho||\sigma) = \sum_i (q_i+r_i)\log(1+r_i/q_i).
\end{eqnarray}
%We can consider two regimes now, one in which $\sigma$ is fixed and $\rho-\sigma$ tends to zero, and
%one in which $\rho-\sigma$ is fixed and $\lambda_{\min}(\sigma)$ tends to zero.
We can develop $S(\rho||\sigma)$ as a Taylor series in the $r_i$, giving
$$
S(\rho||\sigma) = \sum_i \frac{r_i^2}{2q_i} + O(r_i^3).
$$
Hence, in the regime where $\rho-\sigma$ tends to zero and $\sigma$ is otherwise fixed,
the relative entropy exhibits the behaviour of bound (\ref{bound_quad}).
% and (\ref{newb}).

In terms of the matrix derivatives,
this notion can be made more precise as follows. Denote the
bound (\ref{bound_log}) as
$$
    g(\rho||\sigma) = \frac{\trace[(\rho-\sigma)^2]}{\lambda_{\min}(\sigma)}
$$
for states $\rho,\sigma$,
then clearly
$$
\lim_{\varepsilon \rightarrow 0} g(\sigma +\varepsilon \Delta
    ||\sigma)/\varepsilon=0.
$$
On using the integral representation of the second Fr\'echet derivative of
the matrix logarithm \cite{ohya},
\begin{eqnarray*}
\lefteqn{\frac{\partial^2}{\partial \varepsilon^2}\Big|_{\varepsilon=0}
    \log(\sigma + \varepsilon \Delta)} \\
    & =& -2 \int_0^\infty dx (\sigma +x)^{-1} \Delta (\sigma+x)^{-1} \Delta (\sigma+x)^{-1},
\end{eqnarray*}
one obtains
\begin{eqnarray*}
\lefteqn{\frac{\partial^2}{\partial \varepsilon^2}\Big|_{\varepsilon=0} S(\sigma+\varepsilon \Delta||\sigma)} \\
&=& -2 \trace\biggl[\sigma \int_0^\infty dx (\sigma +x)^{-1} \Delta (\sigma+x)^{-1} \Delta (\sigma+x)^{-1}\biggr]\\
& & +2 \trace\left[\Delta\int_0^\infty dx (\sigma+x)^{-1} \Delta (\sigma+x)^{-1}\right].
\end{eqnarray*}
The right hand side is bounded from above by
\begin{eqnarray*}
\lefteqn{\frac{\partial^2}{\partial \varepsilon^2}\Big|_{\varepsilon=0} S(\sigma+\varepsilon \Delta||\sigma)} \nonumber\\
&\leq& \int_0^\infty dx\trace[\Delta(\sigma+x)^{-1}\Delta(\sigma+x)^{-1}],
\end{eqnarray*}
see Ref.\ \cite{lieb,ohya}.
This bound can be written as in Eq.\ (\ref{dec}). Therefore,
one can conclude that
\begin{eqnarray*}
    \frac{\partial^2}{\partial \varepsilon^2}\Big|_{\varepsilon=0}
    S(\sigma+\varepsilon \Delta
    ||\sigma) = \frac{\partial^2}{\partial \varepsilon^2}\Big|_{\varepsilon=0}
    g(\sigma+\varepsilon \Delta
    ||\sigma)
\end{eqnarray*}
holds for all $\Delta$ satisfying
$\trace[\Delta]=0$ if and only if
$\sigma=\identity/d$, where $d$ is the dimension of the underlying
Hilbert space.
These considerations seem to spell doom for any attempt at ``unifying'' the two kinds of upper bounds.
However, below we will see how a certain change of perspective will allow us to get out of the dilemma.
%%%%%%%%%%%%%%%%%%%%%%%%%%%%%%%%%%%%%%%%%%%%%%%%%%%%%%%%%%%%%%%%%%%%%%%%%%%%%%%%%%%%%%%%%%%%%%%%%%%%%%%%

\section{A sharp lower bound in terms of norm distance}

We define $S_{\min}(T)$ with respect to a norm to be 
the smallest relative entropy between two states that have
a distance of exactly $T$ in that norm,
that is
\begin{equation} 
\label{eq:def_smin}
S_{\min}(T) = \min_{\rho,\sigma} \left\{S(\rho||\sigma): |||\rho-\sigma||| = 
T \right\}.
\end{equation}
When one agrees to assign $S(\rho||\sigma)=+\infty$ for non-positive $\rho$,
the definition of $S_{\min}$ can be rephrased as
\begin{equation} \label{eq:def_smin2}
S_{\min}(T) = \min_{\Delta,\sigma} \left\{S(\sigma+\Delta||\sigma): |||\Delta||| = 
T, \trace[\Delta]=0 \right\}.
\end{equation}
Intuitively one would guess that $S_{\min}$ is monotonously increasing with $T$. The following lemma shows
that this is true, but some care is required in proving it.
\begin{lemma}\label{lem:mono}
For $T_1\le T_2$, $S_{\min}(T_1)\le S_{\min}(T_2)$.
\end{lemma}
\textit{Proof.}
Keep $\sigma$ fixed and define
$$
f_\sigma(T) = \min_{\Delta} \left\{S(\sigma+\Delta||\sigma): |||\Delta||| = 
T, \trace[\Delta]=0\right\},
$$
so that $S_{\min}(T) = \min_\sigma f_\sigma(T)$.
Considering $S(\sigma+\Delta||\sigma)$ as a function of $\Delta$, it is convex and minimal in
the origin $\Delta=0$.
Furthermore, for the norm balls 
\begin{equation}\nonumber
\cB(T):=\{\Delta: |||\Delta|||\le T, \trace[\Delta]=0\} 
\end{equation}
we have
\begin{equation}\nonumber
\{0\}=\cB(0)\subseteq\cB(T_1)\subseteq\cB(T_2).
\end{equation}
This is sufficient to prove that
$0=f_\sigma(0)\le f_\sigma(T_1)\le f_\sigma(T_2)$.
Now, since this holds for any $\sigma$, it also holds when minimising over $\sigma$,
and that is just the statement of the Lemma.
\qed

As a direct consequence, a third equivalent definition of $S_{\min}(T)$ is
\begin{equation} \label{eq:def_smin3}
S_{\min}(T) = \min_{\Delta,\sigma} \left \{S(\sigma+\Delta||\sigma): |||\Delta||| 
\ge T, \trace[\Delta]=0 
\right\}.
\end{equation}
We now show that one can restrict oneself to the commutative case.
\begin{lemma}
The minimum in Eq.\ (\ref{eq:def_smin2}) is obtained for $\sigma$ and $\Delta$ commuting.
\end{lemma}
\textit{Proof.}
Fix $\Delta$ and consider a basis in which $\Delta$ is diagonal. Let 
$\rho\longmapsto\diag(\rho)$ be the
completely positive trace-preserving map which, in that basis, sets all off-diagonal elements of $\rho$ equal to zero.
Thus $\diag(\Delta)=\Delta$.
By monotonicity of the relative entropy,
$$
S(\sigma+\Delta||\sigma) \ge S(\diag(\sigma)+\Delta||\diag(\sigma)).
$$
Minimising over all states $\sigma$ then gives
\begin{eqnarray*}
\min_\sigma S(\sigma+\Delta||\sigma) &\ge& \min_\sigma S(\diag(\sigma)+\Delta||\diag(\sigma)) \\
&=& \min_\sigma\left \{S(\sigma+\Delta||\sigma):[\sigma,\Delta]=0\right \}.
\end{eqnarray*}
On the other hand, the states $\sigma$ that commute with $\Delta$ are included in the domain of minimisation
of the left-hand side, hence equality holds.
\qed

\begin{figure}
\includegraphics[width=3.4in]{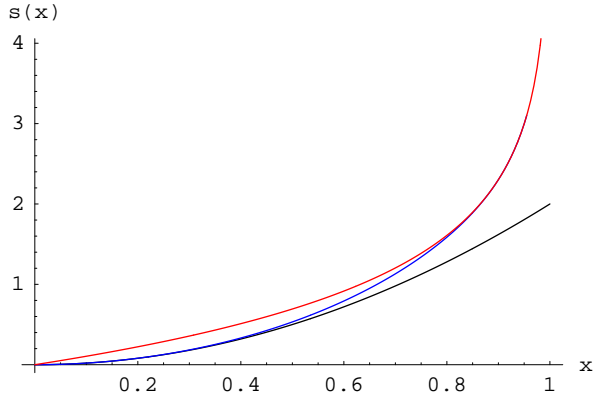}
\caption{\label{fig1}The function $s$ defined in Eq.\  (\ref{eq:sofx})
(middle curve), the lower bound $2x^2$ (lower curve),
and the upper bound $-\log(1-x)$ (upper curve).}
\end{figure}

For later reference we define the auxiliary function
%\begin{definition}
\begin{equation}\label{eq:sofx}
s(x) := \min_{0<r<1-x} S((r+x,1-r-x)||(r,1-r)),
\end{equation}
for $0\le x< 1$.
%\end{definition}
An equivalent expression for this function is given by
\begin{equation}\nonumber
s(x) := \min_{x<r<1} S((r-x,1-r+x)||(r,1-r)).
\end{equation}
The first three non-zero terms in its series expansion around $x=0$ are given by:
\begin{equation}\nonumber
s(x) =
2 x^2 + \frac{4}{9}x^4 + \frac{32}{135}x^6 + O(x^{8}) \label{bound_ohya_better}
\end{equation}
(obtained using a computer algebra package).
Further calculations reveal that some of the higher-order coefficients are negative, the first one being
the coefficient of $x^{62}$.
One can easily prove \cite{csiszar} that
the lowest order expansion $2x^2$ is actually a lower bound. It is,
therefore, the sharpest quadratic lower bound.
For values of $x$ up to $1/2$, 
the error incurred by considering only the lowest order term in (\ref{bound_ohya_better})
is at most 6.5\%. For larger values of $x$, the error increases rapidly.
In fact, when $x$ tends to its maximal value of 1,
$s(x)$ tends to infinity, as can easily be seen from the minimisation expression ($r$ tends to 0);
accordingly, the series expansion diverges.
For values of $x>4/5$, 
$s(x)$ is well approximated by its upper bound
\begin{eqnarray*}
	s(x) &\le & \lim_{r\rightarrow 1-x} S((r+x,1-r-x)||(r,1-r))\nonumber\\
	& = & -\log(1-x).
\end{eqnarray*}
This is illustrated in Figure \ref{fig1}.

Let us now come back to Eq.\ 
(\ref{eq:def_smin2}), with $\sigma$ and $\Delta$ diagonal, and
$|||.|||$ any unitarily invariant norm.
Let $\sigma$ and $\Delta$ have diagonal elements $\sigma_k$ and $\Delta_k$, respectively.
Fixing $\Delta$, we minimise first over $\sigma$.
This is a convex problem and any local minimum is automatically a global minimum
\cite{Convex}. The corresponding Lagrangian 
is
\begin{equation}\nonumber
\cL = \sum_k \sigma_k (1+\Delta_k/\sigma_k)\log(1+\Delta_k/\sigma_k) - \nu
\left(\sum_k\sigma_k-1\right).
\end{equation}
The derivative of the Lagrangian with respect to $\sigma_k$ is
\begin{equation}\label{eq:deriv2}
\frac{\partial\cL}{\partial \sigma_k} = \log(1+\Delta_k/\sigma_k)-\Delta_k/\sigma_k-\nu.
\end{equation}
This must vanish in a critical point, giving the expression
\begin{equation}\nonumber
\log(1+\Delta_k/\sigma_k) = \Delta_k/\sigma_k+\nu. \label{eq:de2}
\end{equation}
Now note that the equation $\log(1+x)-x=b$, for $b<0$ has only two real solutions, one positive and one negative,
and none for $b>0$.
Therefore, for any $k$ $\Delta_k/\sigma_k$ can assume only one of these two possible values.
Let $K$ be an integer between 1 and $d-1$.
Without loss of generality we can set
\begin{equation}\label{eq:ds}
\Delta_k/\sigma_k = \left\{\begin{array}{ll}c_p,& 1\le k\le K,
\\ -c_m,& K<k\le d,\end{array} \right.
\end{equation}
where $c_p$ and $c_m$ are positive numbers, to be determined along with $K$.
The requirement $\sum_k \Delta_k=0$ imposes
$$
c_p \sum_{k=1}^K \sigma_k - c_m \sum_{k=K+1}^d \sigma_k = 0,
$$
which upon defining
\begin{equation}\label{eq:r}
r: = \sum_{k=1}^K \sigma_k,
\end{equation}
turns into
\begin{equation}\label{eq:cc}
c_p r = c_m (1-r) =: c.
\end{equation}
Substituting Eqs.\
(\ref{eq:ds}) and (\ref{eq:r}), the function to be minimised becomes
$$
r (1+c_p)\log(1+c_p) +(1-r)(1-c_m)\log(1-c_m),
$$
which, given Eq.\ 
(\ref{eq:cc}), can be rewritten as
$$
S((r+c,1-r-c)||(r,1-r)).
$$
The one remaining constraint $|||\Delta|||=T$ likewise becomes
$$
|||(c_p\sigma_1,\ldots,c_p\sigma_K,-c_m\sigma_{K+1},\ldots,-c_m\sigma_d)|||=T.
$$
Defining
\begin{eqnarray*}
\tau' &:=& (\sigma_1,\ldots,\sigma_K)/r , \\
\tau'' &:=& (\sigma_{K+1},\ldots,\sigma_d)/(1-r),
\end{eqnarray*}
this turns into
$$
T=|||(c_p r\tau' ; -c_m(1-r)\tau'')|||=c |||(\tau';\tau'')|||,
$$
where we have exploited the homogeneity of a norm.
Note that by their definition, $\tau'$ and $\tau''$ are vectors consisting of positive numbers adding up to 1.

The minimisation itself thus turns into
$$
S_{\min}(T) = \min_{r,\tau',\tau''} S((r+c,1-r-c)||(r,1-r)),
$$
where $c :=T/|||(\tau';\tau'')|||$.
Quite obviously, the minimum over $c$ is obtained for the smallest possible $c$, hence
\begin{eqnarray*}
	S_{\min}(T)  & =& \min_{r} 
	S((r+T/\gamma,1-r-T/\gamma)||(r,1-r))\nonumber\\		
	& = & s(T/\gamma),
\end{eqnarray*}
with
$$
\gamma=\max_{\tau',\tau''} |||(\tau';\tau'')|||.
$$
By convexity of a norm, this maximum is obtained in an extreme point, so
$$
\gamma=|||F|||.
$$
Incidentally, by Lemma \ref{lem:maxt}, this value is also the maximum
$$
\max_{\rho,\sigma}|||\rho-\sigma|||,
$$
over all possible states $\rho$ and $\sigma$, i.e., $\gamma$ is the largest possible value of $T$ for the
given norm.
We have thus proven
\begin{theorem}
For any unitarily invariant norm $|||.|||$, we have the sharp lower bound
\begin{equation}\nonumber
S(\rho||\sigma) \ge s(|||\rho-\sigma|||/|||F|||).
\end{equation}
\end{theorem}

A few remarks are in order at this point:
\begin{enumerate}
\item Within the setting of finite-dimensional systems, this theorem generalises a result of Hiai, Ohya and Tsukada
\cite{hot,ohya} for the trace norm to all unitarily invariant norms.
This paper also uses the technique of getting lower bounds
by projecting on an abelian subalgebra and then 
exploiting the case of a two-dimensional 
support as the worst
case scenario.

\item If we take the Hiai-Ohya-Tsukada result for granted 
and combine it with Lemma \ref{lemma:dom},
we immediately get
\begin{eqnarray*}
	S(\rho||\sigma) &\ge & s(||\rho-\sigma||_1/||F||_1)\nonumber\\
	& \ge & s(|||\rho-\sigma|||/|||F|||).
\end{eqnarray*}
\item The divergence of $s$ at $x=1$ is easily understood.
The largest norm difference between two states occurs for orthogonal pure states, in which case
their relative entropy is infinite.
\end{enumerate}
%%%%%%%%%%%%%%%%%%%%%%%%%%%%%%%%%%%%%%%%%%%%%%%%%%%%%%%%%%%%%%%%%%%%%%%%%%%%%%%%%%%%%%%%%%%%%%%%%%%%%%
\section{Sharp upper bounds in terms of norm distance}
Let now $S_{\max}(T,\beta)$ be the largest relative entropy between $\rho$ and $\sigma$ that have
a normalised distance of exactly $T$ and $\lambda_{\min}(\sigma)=\beta$, 
so let
\begin{equation} \label{eq:def_smax}
S_{\max}(T,\beta) := \max_{\rho,\sigma} \left \{S(\rho||\sigma):
\frac{|||\rho-\sigma|||}{|||F|||} = T, \lambda_{\min}(\sigma)=\beta\right \}.
\end{equation}
The need for the extra parameter $\beta$ arises because for $\beta=0$, $S_{\max}$ is infinite, as can be seen
by taking different pure states for $\rho$ and $\sigma$.
We can rephrase this definition as
\begin{eqnarray} \label{eq:def_smax2}
S_{\max}(T,\beta) &=& \max_{\Delta,\sigma} \Biggl \{S(\sigma+\Delta||\sigma):
\frac{|||\Delta|||}{|||F|||} = T, \trace[\Delta]=0, \nonumber \\
&& \sigma+\Delta\ge0,\lambda_{\min}(\sigma)=\beta\Biggr  \}.
\end{eqnarray}
Because $\Delta$ commutes with the identity matrix, there is a unique common least upper bound on $\beta\identity$
and $-\Delta$, which we will denote by $\max(\beta\identity,-\Delta)$. In the eigenbasis of $\Delta$, this is a diagonal
matrix with diagonal elements $\max(\beta,-\Delta_{i})$.
The constraints $\sigma\ge\beta$ and $\sigma+\Delta\ge 0$ can therefore be combined into the single constraint
\begin{equation}\label{eq:sigcon}
\sigma \ge \max(\beta\identity,-\Delta).
\end{equation}
The extremal $\sigma$ obeying this constraint are
\begin{equation}\nonumber
\sigma = \max(\beta\identity,-\Delta) + \eta \ket{\psi}\bra{\psi},
\end{equation}
where $\psi$ is any state vector, 
and
\begin{equation}\nonumber
\eta := 1-\trace[\max(\beta\identity,-\Delta)].
\end{equation}
Therefore, the constrained maximisation over $\sigma$ can be replaced by an unconstrained maximisation over all
pure states of the function
\begin{eqnarray}
S(\Delta+\max(\beta\identity,-\Delta) + \eta \ket{\psi}\bra{\psi}\,\, || \nonumber \\
\qquad \max(\beta\identity,-\Delta) + \eta \ket{\psi}\bra{\psi}). \label{eq:maximand}
\end{eqnarray}
Of course, all of this puts constraints on $\Delta$ as well. Indeed, in order that states $\sigma$
obeying (\ref{eq:sigcon}) exist,
$\max(\beta\identity,-\Delta)$ must obey the condition
\begin{equation}\nonumber
\trace[\max(\beta\identity,-\Delta)]\le 1.
\end{equation}
We now have to distinguish between two cases: the case $d=2$, and the case $d>2$.
%%%%%%%%%%%%%%%%%%%%%%%%%%%%%%
\subsection{The case $d=2$}
For the $d=2$ case, the maximisation over $\Delta$ is trivial. In its eigenbasis,
$\Delta$ is a multiple of $\diag(1,-1)=F$. Hence, fixing the eigenbasis of $\Delta$ (which we can do because of unitary
invariance of the relative entropy), and fixing 
\begin{equation}\nonumber
|||\Delta|||/|||F|||=T, 
\end{equation}
actually leaves just one possibility
for $\Delta$, namely $\Delta=TF$.
The term $\max(\beta\identity,-\Delta)$ leads to two cases: $T\le\beta$ and $T>\beta$.

%%%%%%%%%%%%%%%%%%%%%%%%%%%\subsubsection{The case $T\le \beta$}
The condition $T\le\beta$ implies, by Lemma \ref{lemma:dom}, that $||\Delta||_\infty\le\beta$ and, hence,
\begin{eqnarray*}
\max(\beta\identity,-\Delta) &=& \diag(\beta,\beta) ,\\
\eta &=& 1-2\beta.
\end{eqnarray*}
The remaining maximisation of (\ref{eq:maximand}) is therefore given by
\begin{eqnarray}
\max_\psi S(\diag(\beta+T,\beta-T)+(1-2\beta)\ket{\psi}\bra{\psi} \, || \nonumber\\
\qquad \diag(\beta,\beta)+(1-2\beta)\ket{\psi}\bra{\psi}).\label{eq:maxi1a}
\end{eqnarray}
Positivity of $\eta$ requires $\beta\le 1/2$.
By unitary invariance of the relative entropy, and invariance of diagonal states under
diagonal unitaries (phase factors),
we can restrict ourselves to vectors 
$\psi$ of the form $\psi=(\cos\alpha,\sin\alpha)^T$.

\begin{lemma}\label{lem:convex}
For a state vector $\psi=(\cos\alpha,\sin\alpha)^T$, the function to be maximised
in (\ref{eq:maxi1a}) is convex in $\cos(2\alpha)$.
\end{lemma}
\textit{Proof.}
Let $D_1$ be the determinant of the first argument.
It is linear in $t:=\cos(2\alpha)$:
$$
D_1 = \beta^2-T^2+(1-2\beta)(\beta-T t).
$$
After some 
basic algebra involving eigensystem decompositions of the states,
the function to be maximised in (\ref{eq:maxi1a}) is found to be
given by
\begin{eqnarray*}
f(x) &:=& ((1-x)\log(1-x)+(1+x)\log(1+x))/2  \\
&+& (-1+2\beta-2T t)(\log(1-\beta)-\log\beta)/2 \\
&-& (\log(4-4\beta)+\log\beta)/2,
\end{eqnarray*}
where $x=(1-4D_1)^{1/2}$.
We will now show that this function is convex in $t$.
Since the second and third terms are linear in $t$, we only need to show convexity for the first term.
The series expansion of the first term is
$$
((1-x)\log(1-x)+(1+x)\log(1+x))/2 = \sum_{k=1}^\infty \frac{x^{2k}}{2k(2k-1)}.
$$
Every term in the expansion is a positive power of $x^2$ with positive coefficient
and is therefore convex in $x^2$, which itself is linear in
$t$. The sum is therefore also convex in $t$.
\qed

By the above Lemma, the maximum of the maximisation over $\psi$ is obtained
for extremal values of $t$, that is: either $\psi=(1,0)^T$ or $\psi=(0,1)^T$.
Evaluation of the maximum is now straightforward and it can be checked that the choice
$\psi=(1,0)^T$ always yields the largest value of the relative entropy.

%%%%%%%%%%%%%%%%%%%%%%%%%\subsubsection{The case $T>\beta$}

We will now more specifically look at the case where $T>\beta$.
In this case, we get
%
%When $T>\beta$, we obtain
\begin{eqnarray*}
\max(\beta\identity,-\Delta) &=& \diag(\beta,T), \\
\eta &=& 1-\beta-T,
\end{eqnarray*}
and the remaining maximisation of (\ref{eq:maximand}) is given by
\begin{eqnarray}
\max_\psi S(\diag(\beta+T,0)+(1-\beta-T)\ket{\psi}\bra{\psi} \, || \nonumber\\
\qquad \diag(\beta,T)+(1-\beta-T)\ket{\psi}\bra{\psi}).\label{eq:maxi1}
\end{eqnarray}
Positivity of $\eta$ requires $\beta\le 1/2$ and $T\le 1-\beta$.
Again, we can restrict ourselves to states $\psi=(\cos\alpha,\sin\alpha)^T$.
We also have the equivalent of Lemma 
\ref{lem:convex}, which needs more work in this case:
\begin{lemma}
For a state vector $\psi=(\cos\alpha,\sin\alpha)^T$, the function to be maximised
in (\ref{eq:maxi1}) is convex in $\cos(2\alpha)$.
\end{lemma}
\textit{Proof.}
Let $D_1$ and $D_2$ be the determinant of the first and second argument, respectively.
Both are linear in $t:=\cos(2\alpha)$:
\begin{eqnarray*}
D_1 &=& (1-\beta-T)(\beta+T)(1-t)/2 \\
D_2 &=& ((\beta+T-\beta^2-T^2)+(1-\beta-T)(T-\beta)t)/2.
\end{eqnarray*}
In the $(D_1,D_2)$-plane, this describes a line segment with gradient
$$
K:= -\frac{T-\beta}{T+\beta},
$$
which lies in the interval $[-1,0]$.

Again, after some 
basic algebra,
the function to be maximised in (\ref{eq:maxi1}) is 
identified to be 
$f((1-4D_1)^{1/2}, (1-4D_2)^{1/2})$, where
\begin{eqnarray*}
f(x,y) &:=& ((1-x)\log(1-x)+(1+x)\log(1+x))/2  \\
&+& ((x^2+y^2-2y-4T^2)\log(1-y)  \\
&-& (x^2+y^2+2y-4T^2)\log(1+y))/4 y.
\end{eqnarray*}
We will now show that $f((1-4D_1)^{1/2},
(1-4D_2)^{1/2})$ is convex in $t$.
First, note that 
\begin{eqnarray*}
	f(x,y)=f_0(x,y)+T^2 f_1(y).
\end{eqnarray*}
The term $f_1(y)$ is itself convex in $t$: its series expansion is
$$
f_1(y) = (\log(1+y)-\log(1-y))/y = 2\sum_{k=0}^\infty \frac{y^{2k}}{2k+1},
$$
which by the positivity of all its coefficients is convex in $y^2$, and $y^2$ is linear in $t$.

The other term, $f_0(x,y)$ is given by a sum of three terms
\begin{eqnarray*}
f_0(x,y) &=& \frac{1}{2}((1-x)\log(1-x)+(1+x)\log(1+x)) \\
&+&  \frac{1}{4}((y-2)\log(1-y)-(y+2)\log(1+y)) \\
&-&  \frac{x^2}{4}(\log(1+y)-\log(1-y))/y.
\end{eqnarray*}
Replacing each of the three terms by its series expansion yields
\begin{eqnarray*}
f_0(x,y) &=& \sum_{k=1}^\infty \frac{x^{2k}}{2k(2k-1)} \\
&+&  \sum_{k=1}^\infty (k-1)\frac{y^{2k}}{2k(2k-1)} \\
&-& \frac{x^2}{2}\sum_{k=0}^\infty \frac{y^{2k}}{2k+1}.
\end{eqnarray*}
To show that this function is convex in $t$, we will evaluate it along the curve
\begin{eqnarray*}
x^2 &=& u+p \\
y^2 &=& v+Kp,
\end{eqnarray*}
with gradient $K$ between 0 and $-1$,
and $u$ and $v$ lying in the interval $[0,1]$,
and check positivity of its second derivative with respect to
 $p$ at $p=0$:
\begin{eqnarray*}
	&&{\left.
	\frac{\partial^2}{\partial p^2}\right |_{p=0} 
	f_0(x,y)} 
	=\sum_{k=2}^\infty \frac{k-1}{2k-1}u^{k-2} \\
	&&+(k-1)\left(K\frac{(k-1)K-2}{2k-1} - 
	K^2 \frac{k}{2k+1}u \right)v^{k-2}.
\end{eqnarray*}
The coefficient of $u^{k-2}$ is clearly positive, hence
the derivative is positive if the coefficient of $v^{k-2}$ is positive for all allowed values of $u$ and $K$.
The worst case occurs for $u=1$, yielding a coefficient
$$
K\frac{(k-1)K-2}{2k-1} - K^2 \frac{k}{2k+1} = \frac{-K (2 + 4 k + K)}{(2k-1)(2k+1)}.
$$
For values of $K$ between 0 and $-1$, this is indeed positive.
\qed

By the above Lemma, the maximum of the maximisation over $\psi$ is obtained
for extremal values of $t$, that is: either $\psi=(1,0)^T$ or $\psi=(0,1)^T$.
Evaluation of the maximum is again straightforward, and calculations show that
sometimes $\psi=(1,0)^T$ yields the larger value, and sometimes $\psi=(0,1)^T$.
In this way we have obtained the upper bounds:
\begin{theorem}\label{th:ub2b}
Let $\Delta=\rho-\sigma$, $T=|||\Delta|||/|||F|||$ and $\beta=\lambda_{\min}(\sigma)$.
For $d=2$, and $T\le\beta$,
\begin{eqnarray}
S(\rho||\sigma) &\le& (T+1-\beta)\log\frac{T+1-\beta}{1-\beta} \nonumber \\
&+& (\beta-T)\log(1-T/\beta) \label{bound_quad2}.
\end{eqnarray}
For $d=2$, and $T>\beta$,
\begin{eqnarray}
S(\rho||\sigma) &\le& \max(-\log(1-T) , \nonumber\\
&& (\beta+T)\log(1+T/\beta)+\nonumber\\
&& (1-\beta-T)\log(1-T/(1-\beta))).
\end{eqnarray}
\end{theorem}

It is interesting to study the behaviour of the bound in the
case of  large 
$\beta$. More specifically,
an approximation for bound (\ref{bound_quad2}), valid for $T \ll \beta$, is
\begin{eqnarray}
S(\rho||\sigma) &\le& \sum_{k=2}^\infty \frac{T^k}{k(k-1)}
\left(\frac{1}{\beta^k}-\frac{(-1)^k}{(1-\beta)^k}\right) \nonumber \\
&\approx& \frac{T^2}{2\beta(1-\beta)}, \label{bound_quad2_approx}
\end{eqnarray}

\begin{figure*}
\begin{tabular}{cc}
\includegraphics[width=3.4in]{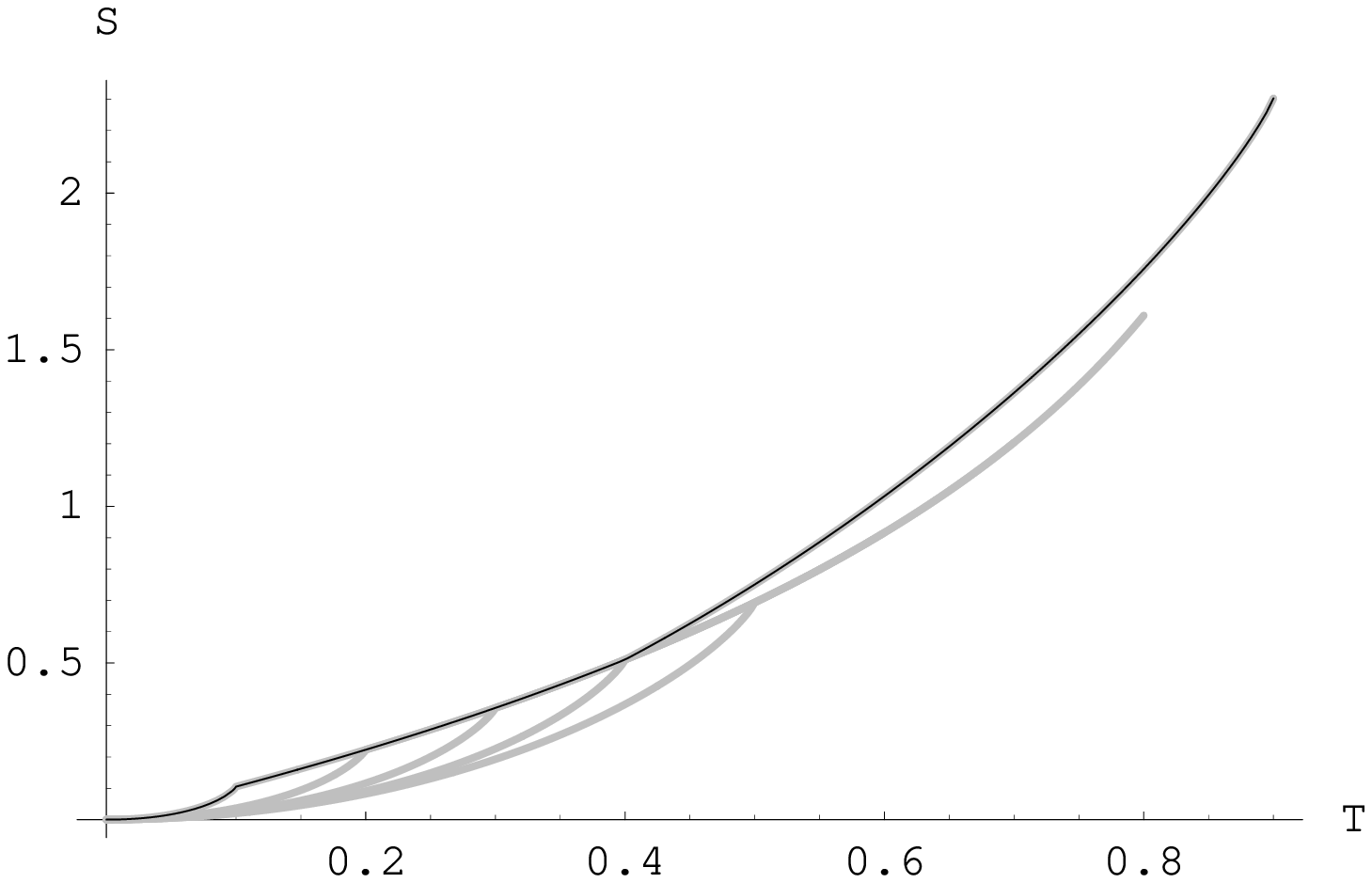} & \includegraphics[width=3.4in]{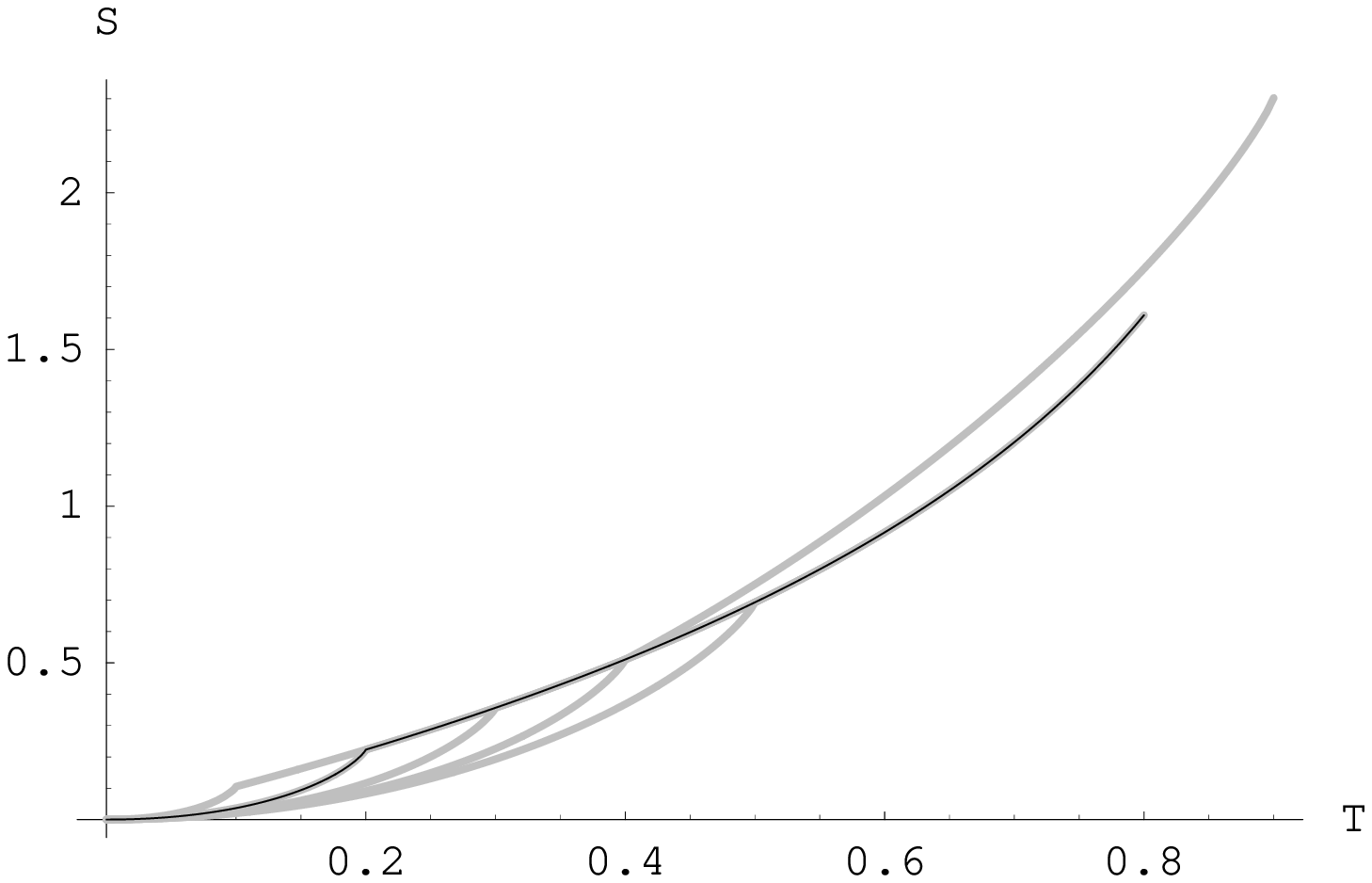} \\
(a) & (b) \\
\includegraphics[width=3.4in]{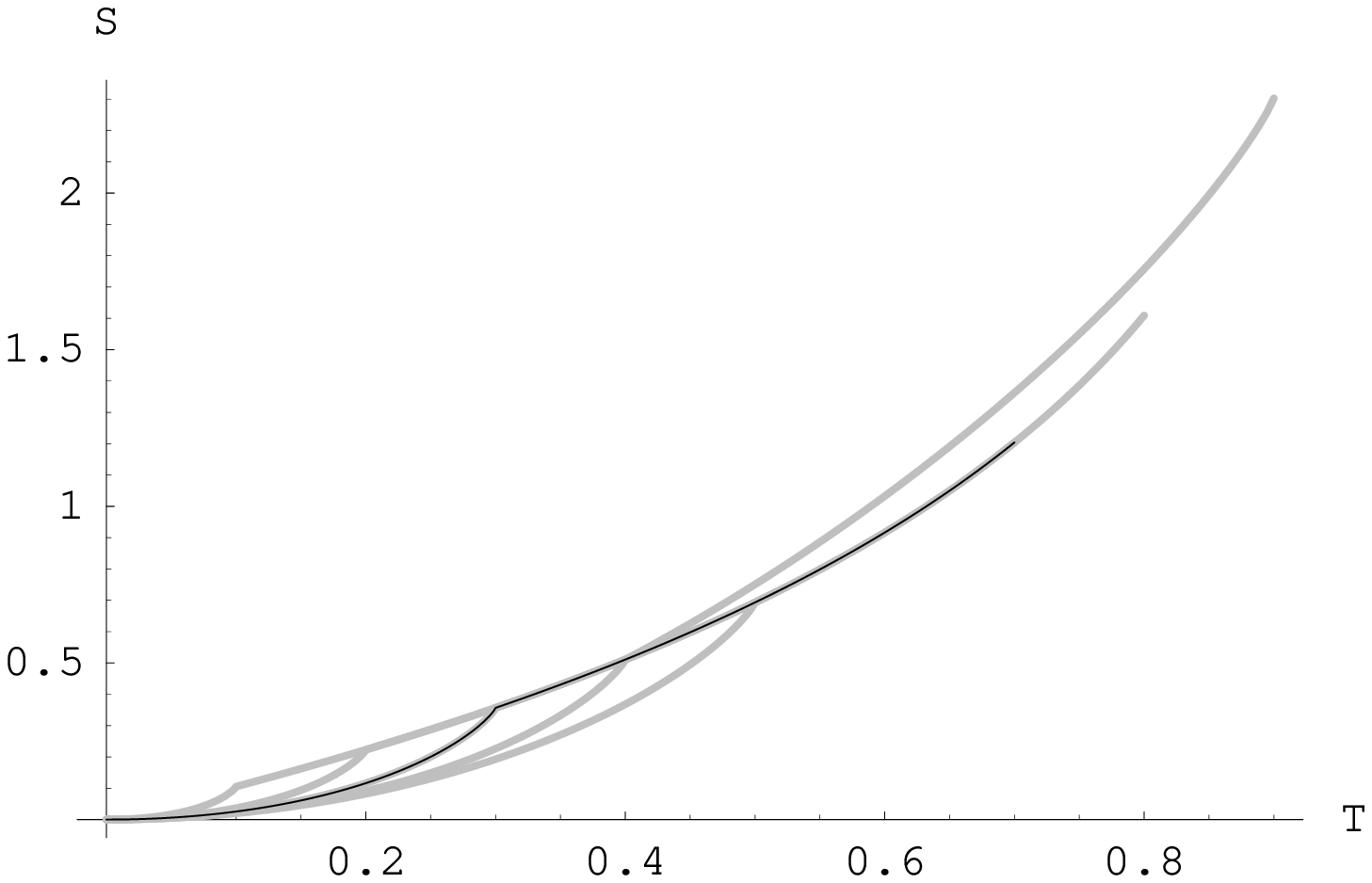} & \includegraphics[width=3.4in]{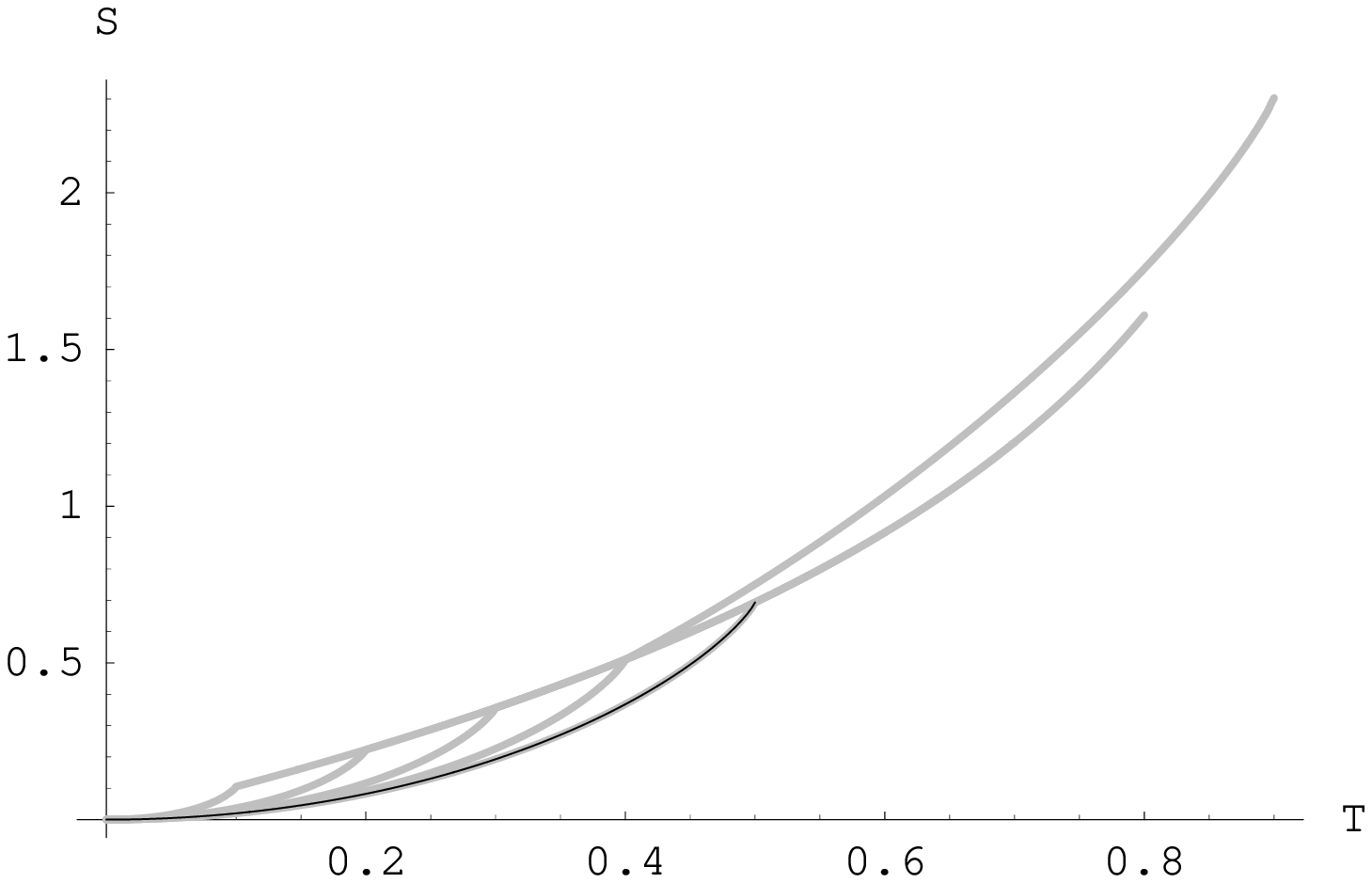} \\
(c) & (d)
\end{tabular}
\caption{\label{fig2}Upper bounds of Theorem \ref{th:ub2b}
on $S=S(\rho||\sigma)$ vs.\ the rescaled norm distance $T=|||\rho-\sigma|||/|||F|||$,
for $d=2$, and for values of smallest eigenvalue of $\sigma$ (a) $\beta=0.1$, (b) $0.2$, (c) $0.3$, and (d) $0.5$.
The two regimes $T\le\beta$ and $\beta\le T\le 1-\beta$ can be clearly identified.
For ease of 
comparison, each curve is shown superimposed on the curves for
$\beta=0.1$, $0.2$, $0.3$, $0.4$ and $0.5$ (in grey).}
\end{figure*}

Figure \ref{fig2} illustrates the combined upper bounds of Theorem \ref{th:ub2b}
($d=2$) for various values of $\beta$.
%%%%%%%%%%%%%%%%%%%%%%%%%%%%%%
\subsection{The case $d>2$}
In case $d$ is larger than 2, it is not clear how to proceed in the most general setting, for general UI norms,
as the maximisation over $\Delta$ must explicitly be performed.
In the following, we will restrict ourselves to using the trace norm, which is in some sense the most important
one anyway. That is, the requirements on $\Delta$ are
\begin{eqnarray}
||\Delta||_1 &=& 2T, \label{eq:delta1} \\
\trace[\Delta] &=& 0 ,\label{eq:delta2}  \\
\trace[\max(\beta\identity,-\Delta)] &\le& 1. \label{eq:delta3}
\end{eqnarray}
The following very simple Lemma will prove to be a powerful tool.
\begin{lemma}\label{lem:sabc}
For all $A$, $B$, and $C$, positive semi-definite operators,
$$
S(A+C||B+C) \le S(A||B).
$$
\end{lemma}
\textit{Proof.}
First note that for any $a>0$,
\begin{eqnarray*}
S(aA||aB) &=& \trace[aA(\log(aA)-\log(aB))] \\
&=& a S(A||B).
\end{eqnarray*}
This, together with joint convexity of the relative entropy in its arguments (which
need not be normalised to trace 1), leads to
\begin{eqnarray*}
S(A+C||B+C) &=& 2 S(\frac{A+C}{2} || \frac{B+C}{2}) \\
&\le& S(A||B) + S(C||C) \\
&=& S(A||B).
\end{eqnarray*}
\qed

The Lemma immediately yields an upper bound on (\ref{eq:maximand}):
%\begin{eqnarray}
%&&S(\Delta+\max(\beta\identity,-\Delta) + \eta \ket{\psi}\bra{\psi}\,\, || \nonumber \\
%&&\qquad \max(\beta\identity,-\Delta) + \eta \ket{\psi}\bra{\psi}) \nonumber \\
%&\le& S(\Delta+\max(\beta\identity,-\Delta)\,||\, \max(\beta\identity,-\Delta)) \nonumber \\
%&=& S((\Delta+\beta\identity)_+ \,||\, \beta\identity + (\Delta+\beta\identity)_-).\label{eq:s11}
%\eea
letting
$$
\sigma:=\max(\beta\identity,-\Delta) + \eta \ket{\psi}\bra{\psi},
$$
such that we obtain
\begin{eqnarray}
S(\Delta+\sigma \,||\,\sigma)
&\le& S(\Delta+\max(\beta\identity,-\Delta)\,||\, \max(\beta\identity,-\Delta)) \nonumber \\
&=& S((\Delta+\beta\identity)_+ \,||\, \beta\identity + (\Delta+\beta\identity)_-).\label{eq:s11}
\end{eqnarray}
To continue, we consider two cases.

%%%%%%%%%%%%%%%%%%%%%%%%%%%% T<b

{\it Case 1 :}
When $T\le\beta$, the requirement (\ref{eq:delta3}) is automatically satisfied,
and $\max(\beta\identity,-\Delta)=\beta\identity$.
Let $\Delta_+$ and $\Delta_-$ be the positive and negative part of $\Delta$, respectively.
That is, $\Delta = \Delta_+ - \Delta_-$, with $\Delta_+$ and $\Delta_-$ non-negative and orthogonal.
Because we are using the trace norm we can rewrite the conditions on $\Delta$ as
\begin{eqnarray*}
||\Delta||_1 &=& \trace[\Delta_+] + \trace[\Delta_-] = 2T ,\\
\trace[\Delta] &=& \trace[\Delta_+ ]- \trace[\Delta_- ]= 0,
\end{eqnarray*}
hence
$$
\trace[\Delta_+ ]= \trace[\Delta_-]= T.
$$
By Lemma \ref{lem:sabc}, (\ref{eq:maximand}) is upper bounded by $S(\Delta+\beta\identity||\beta\identity)$.
By convexity, its maximum over $\Delta_+$, $\Delta_-\ge0$, with $\trace[\Delta_+ ]= 
\trace[\Delta_-] = T$, is obtained
in $\Delta_+$ and $\Delta_-$ of rank 1, giving as upper bound
$$
S(\Delta + \sigma||\sigma) \leq
(\beta+T)\log\frac{\beta+T}{\beta}
+(\beta-T)\log\frac{\beta-T}{\beta}.
$$
The upper bound can be achieved in dimensions $d\ge3$
for all values of $T\le\beta$ by setting $\Delta=TF$ and $\psi=e^3$.

%%%%%%%%%%%%%%%%%%%%%%%%%%%%%% T>b

{\it Case 2:}
In the other case, when $T>\beta$, we have to deal with condition (\ref{eq:delta3}).
To do that we split $\Delta$ into three non-negative parts,
\begin{equation}\nonumber
	\Delta = \Delta_+ -\Delta_0-\Delta_-, 
\end{equation}
with $\Delta_+$, $\Delta_0$ and $\Delta_-$,
operating on orthogonal subspaces $V_+$, $V_0$ and $V_-$, respectively,
with
\begin{eqnarray*}
                  \phantom{-}\Delta_+  &\ge& 0, \\
                0 &\ge& -\Delta_0\, \ge -\beta\identity_0, \\
-\beta\identity_- &\ge& -\Delta_-.
\end{eqnarray*}
We denote the projectors on these subspaces by $\identity_+$, $\identity_0$, and $\identity_-$.
Then
$$
(\Delta+\beta)_+ = \Delta_+ - \Delta_0 + \beta\identity_{+0},
$$
where $\identity_{+0}:=\identity_++\identity_0$.
The conditions on $\Delta$, $\trace[\Delta]=0$ and $\trace[ |\Delta| ]=2T$ translate to
$$
\trace[\Delta_+ ]= \trace[\Delta_0]+\trace[\Delta_- ]= T.
$$
Due to the orthogonality of positive and negative part, (\ref{eq:s11}) can be simplified to
$
S((\Delta+\beta\identity)_+ \,||\, \beta\identity_{+0})$.
After subtracting $\beta\identity_0-\Delta_0$ from both arguments,
we get
$$
S(\Delta_++\beta\identity_+ || \beta\identity_++\Delta_0),
$$
which is an upper bound on (\ref{eq:s11}), by Lemma \ref{lem:sabc}.
Ignoring condition (\ref{eq:delta3}) on $\Delta$, we get
$$
S_{\max} \le \max_{\Delta_+\ge 0 \atop \trace[\Delta_+ ]= T}
S(\Delta_++\beta\identity_+ || \beta\identity_+).
$$
By convexity, the maximum is obtained for $\Delta_+$ rank 1, giving
the upper bound
$$
S_{\max}\le(T+\beta)\log((T+\beta)/\beta).
$$

To see that this bound is sharp for (almost) any value of $T$,
consider the two states
\begin{eqnarray*}
\rho &=& \diag(T+\beta,0,0^{\times J},\beta^{\times K},\beta+\eta) , \\
\sigma &=& \diag(\beta,T-J\beta,\beta^{\times J},\beta^{\times 
K},\beta+\eta) ,\\
\eta &:=& 1-T-(d-1-J)\beta.
\end{eqnarray*}
Here, $J$ is an integer between 0 and $d-3$ and $k=d-3-J$.
Conditions on $J$ are $J\beta\le T$ (so that $\sigma\ge0$) and $T\le 1-(d-1-J)\beta$
(so that $\eta\ge0$).
This choice of states can thus be obtained for $\beta\le T\le 1-2\beta$.
It can be seen that $||\rho-\sigma||_1=2T$ and 
\begin{equation}\nonumber
	S(\rho||\sigma)=(T+\beta)\log((T+\beta)/\beta).
\end{equation}
%For $1-2\beta<T\le 1-\beta$ we haven't found an example for sharpness, and it may well
%be that in this regime $S_{\max} = (T+\beta)\log((T+\beta)/\beta)+(1-T-\beta)\log((1-T-\beta)/\beta)$.
%
The result of the foregoing can be subsumed into the following theorem.

\begin{figure}
\includegraphics[width=3.4in]{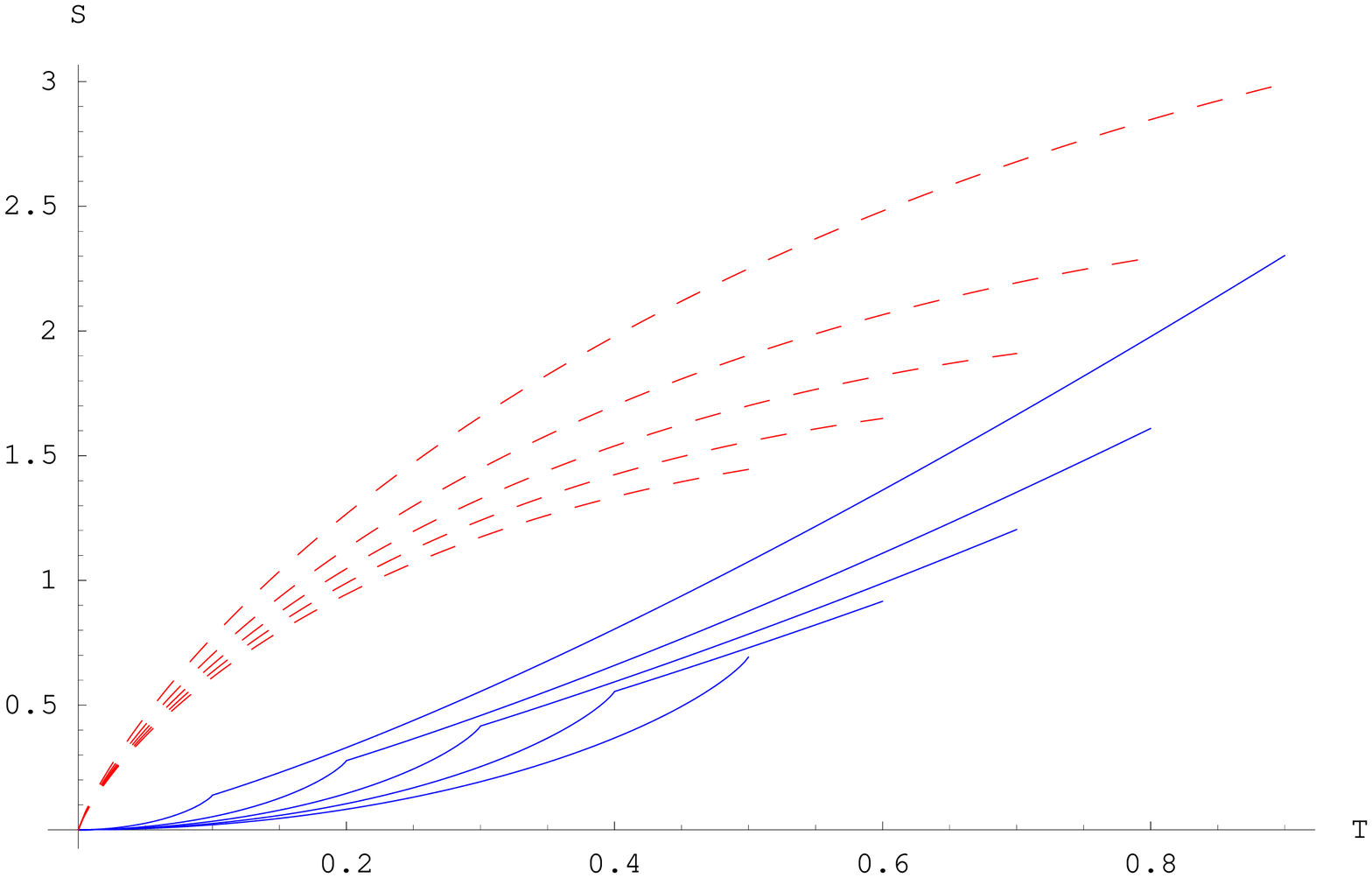}
\caption{\label{fig3}Comparison between upper bounds (\ref{bound_log}) and (\ref{bound_quad2_bis})-(\ref{bound_quad3})
on $S=S(\rho||\sigma)$ vs.\ the trace norm distance $T=||\rho-\sigma||_1/2$,
for various values of $\beta$, the smallest eigenvalue of $\sigma$.
The upper set of dashed curves depict bound (\ref{bound_log}) (with $d=3$)
for $\beta=0.1$ (lower curve), $0.2$, $0.3$, $0.4$ and $0.5$ (upper curve).
The lower set of full line curves depict bounds (\ref{bound_quad2_bis})-(\ref{bound_quad3})
for $\beta=0.1$ (upper curve), $0.2$, $0.3$, $0.4$ and $0.5$ (lower curve).
The two regimes $T\le\beta$ and $\beta\le T\le 1-\beta$ can be clearly seen.
}
\end{figure}

\begin{theorem}
Let $\Delta=\rho-\sigma$, $T=||\Delta||_1/2$ and $\beta=\lambda_{\min}(\sigma)$.
If $T \le \beta$ then
\begin{equation}\nonumber
S(\rho||\sigma) \le (\beta+T)\log\frac{\beta+T}{\beta}
+(\beta-T)\log\frac{\beta-T}{\beta}, \label{bound_quad2_bis}
\end{equation}
and this upper bound is sharp when $d>2$.
If $\beta\le T \le 1-\beta$ then
\begin{equation}\nonumber
S(\rho||\sigma) \le (\beta+T)\log\frac{\beta+T}{\beta}. \label{bound_quad3}
\end{equation}
When $d>2$, this bound is sharp for (at least) $\beta\le T\le 1-2\beta$.
\end{theorem}
Figure \ref{fig3} illustrates these bounds and shows their superiority to the previously obtained bound
(\ref{bound_log}).

Again, it is interesting to look at the bound for large $\beta$.
An approximation for bound (\ref{bound_quad2_bis}), valid for $T \ll \beta$, is
given by
\begin{eqnarray}
S(\rho||\sigma)  \le \sum_{k=1}^\infty \frac{T^{2k}}{k(2k-1) \beta^{2k-1}}
  \approx   
\frac{T^2}{\beta}. \label{bound_quad2_bis_approx}
\end{eqnarray}

%%%%%%%%%%%%%%%%%%%%%%%%%%%%%%%%%%%%%%%%%%%%%%%%%%%%%%%%%%%%%%%%%%%%%%%%%%%%%%%%%%%%%%%%%%%%%%%%%

\section{Application to state approximation}

In the following paragraph we will give an application of our bounds
to state approximation. Consider a state $\rho$ on a Hilbert space $\cH$, and 
a sequence $\{\sigma_n\}_n$
where $\sigma_n$ is a state on $\cH^{\otimes n}$. As before,
the sequence is said to asymptotically approximate $\rho$ if for
$n$ tending to infinity, $
\| \sigma_n-\rho^{\otimes n}\|_1= \trace|\sigma_n-\rho^{\otimes n}|$
tends to zero.
Let us define $T_n$ as
$$
T_n := \trace|\rho^{\otimes n}-\sigma_n|/2.
$$
Because of the lower bound (\ref{bound_ohya}), we get
$$
S_n:=S(\rho^{\otimes n}||\sigma_n) \ge 2 T_n^2,
$$
and this bound is sharp.
Hence, $T_n$ goes to zero if $S_n$ does.

On the other hand, $T_n$ going to zero does not necessarily imply $S_n$ going to zero.
Indeed, $S_n$ can be infinite for any finite value of $n$ when $\rho^{\otimes n}$ is not restricted to the
range of $\sigma_n$. In particular, the relative entropy distance between two pure states is infinite unless
the states are identical. At first sight, this seems to render the relative entropy useless as a distance measure.
Nevertheless, sense can be made of it by imposing an additional requirement that the range of $\sigma_n$ must
contain the range of $\rho^{\otimes n}$.
Let us then restrict $\sigma_n$ to the range of $\rho^{\otimes n}$, as the relative entropy
only depends on that part of $\sigma_n$.
Letting $d$ be the rank of $\rho$, the dimension of the range of $\rho^{\otimes n}$ is $d^n$.
Let $\beta_n$ be the smallest non-zero eigenvalue of $\sigma_n$ on that range;
$\beta_n$ is at most $1/d^n$.

The behaviour of the relative entropy then very much depends on the relation between $\beta_n$ and $T_n$.
Since $\beta_n$ decreases at least exponentially, we only need to consider the case $T_n\ge\beta_n$, and
use the bound (\ref{bound_quad3})
$$
S_n \le (\beta_n+T_n)\log \biggl (1+\frac{T_n}{\beta_n}\biggr).
$$
In the worst-case behaviour of $T_n$ ($T_n/\beta_n$ tending to infinity) the bound can be approximated by
\begin{eqnarray*}
	S_n & \le &  T_n\log\frac{T_n}{\beta_n} = T_n(\log 
	T_n-\log\beta_n)\nonumber\\
	& \approx & T_n|\log\beta_n|.
\end{eqnarray*}
To guarantee convergence of $S_n$ we therefore need $T_n$ to converge to 0 at least as fast as
$1/|\log\beta_n|$, which in the best case goes as $1/n$.
Note that bound (\ref{bound_log}) yields the same requirement, but as this bound is not a sharp one
it could have been too strong a requirement.
This gives us the subsequent theorem.

\begin{theorem}
Consider a state $\rho$ on a finite-dimensional Hilbert space $\cH$
and a sequence $\{\sigma_n\}_n$ of states $\sigma_n$ on $\cH^{\otimes n}$.
The sequence $\{\sigma_n\}_n$ asymptotically approximates $\rho$ in the trace norm,
if 
\begin{equation}\nonumber	
	\lim_{n\rightarrow\infty}
	S(\rho^{\otimes n}||\sigma_n)=0.
\end{equation}
Conversely, if the range of $\sigma_n$ includes the range of $\rho^{\otimes n}$
and $||\rho^{\otimes n}-\sigma_n||_1$ converges to zero faster than $1/|\log\beta_n|$,
where $\beta_n$ is the minimal eigenvalue of $\sigma_n$ restricted to the range of $\rho^{\otimes n}$,
then $\lim_{n\rightarrow\infty} S(\rho^{\otimes n}||\sigma_n)=0$.
\end{theorem}

%%%%%%%%%%%%%%%%%%%%%%%%%%%%%%%%%%%%%%%%%%%%%%%%%%%%%%%%%%%%%%
\section{Summary}

In this paper, we have discussed several lower and upper bounds on the
relative entropy functional, thereby sharpening the notion of
continuity of the relative entropy for states which are
close to each other in the trace norm sense.

The main results are the sharp lower bound from Theorem 4, and the sharp upper bounds of
Theorems 5 ($d=2$) and 6 ($d>2$). Theorems 4 and 5 give the relation between relative entropy
and norm distances based on any unitarily invariant norm, while Theorem 6 holds only for the
trace norm distance. These results have been obtained employing methods from 
optimisation theory.

\begin{acknowledgments}

This work was supported by the Alexander-von-Humboldt Foundation,
the European Commission (EQUIP, IST-1999-11053), the DFG 
(Schwerpunktprogramm QIV), the EPSRC QIP-IRC, 
and the European Research Councils 
(EURYI).

\end{acknowledgments}
%%%%%%%%%%%%%%%%%%%%%%%%%%%%%%%%
%------------------------------------------------------------- BIBLIOGRAPHY

%%%%%%%%%%%%%%%%%%%%%%%%%%%%%%%%%%%%%%%%%%%%%%%%%%%%%%%%%%%%%%%%%%%

\end{document}